\documentclass[twocolumn,english,aps,prl,showpacs]{revtex4-1}
\usepackage[T1]{fontenc}
\usepackage{amsmath,graphicx,amssymb,epsfig,babel,dsfont}
\usepackage{color}

\renewcommand{\Im}{{\rm Im}}
\newcommand{\Tr}{{\rm Tr}}
\newcommand{\rd}{{\rm d}}
\newcommand{\kb}{k_{\rm B}}

\begin{document}

\title{Thermal photon drag in many-body systems}

\author{P. Ben-Abdallah}
\email{pba@institutoptique.fr} 
\affiliation{Laboratoire Charles Fabry,UMR 8501, Institut d'Optique, CNRS, Universit\'{e} Paris-Sud 11,
2, Avenue Augustin Fresnel, 91127 Palaiseau Cedex, France.}

\date{\today}

\begin{abstract}
We demonstrate the existence of a thermal analog of Coulomb drag  in many-body systems which is driven by thermal photons. We show that this frictional effect can either be positive or negative depending on the separation distances within the system. Also we highlight that the persistent heat currents flowing in  non-reciprocal systems at equilibrium are subject to this effect and the latter can even amplify these flows.
\end{abstract}

\maketitle

Spatially separated electric conductors are coupled through the interactions of free charge carriers  provided their separation distance is small enough in comparison with the range of Coulombic interactions. Hence, when a current flow in one also called drive conductor it induces a current by Coulomb drag~\cite{Pogrebinskii,Price,Narozhny,Bhandari} in a second (passive) conductor placed close to it (Fig.~1-a). 
In this Letter we demonstrate the existence of a thermal analog of this effect in many body systems in which heat exchanges are mediated by thermal photons. In these systems the local thermal fluctuations within each body give rise to oscillations of partial charges which, in turn, radiate a time-dependent electric field in their surrounding environement. This leads through the many-body interactions (i.e. multiscattering combined to spontaneous absorption/emission of thermal  photons) to an exchange of a net heat flux between the different regions of the system which have different temperatures~\cite{PBAEtAl2011,Riccardo,Nikbakht,Incardone,splitter,PRL_superdiff}. After introducing the concept of thermal photons drag resistance we investigate the drag effect  in a four-terminal system composed by two parallel pairs of nanoparticles when a temperature difference is held along one of these pairs (drive) while the second pair (passive) is left free to relax. We show that the sign of the temperature difference induced in the passive pair of nanoparticles can be either positive or negative depending on the strength and the nature of interactions between the particles. Finally we show that the heat supercurrents in non-reciprocal many body systems are also subject to this  drag effect and the latter can amplify or inhibitate these persistent heat currents.

\begin{figure}
\centering
\includegraphics[scale=0.3]{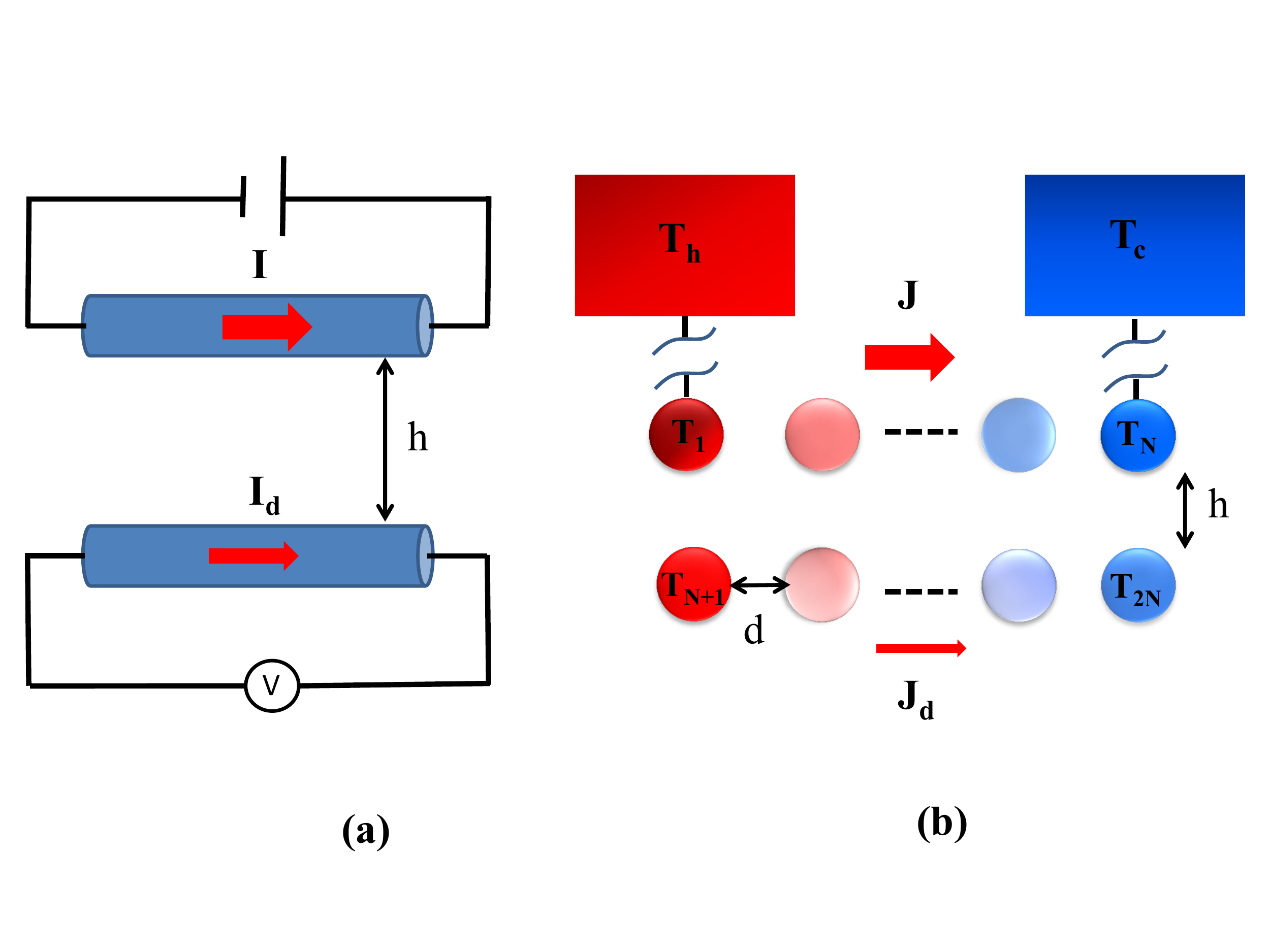}
\caption{(a) Coulomb drag: a drag electric current  $I_d$ in a passive conducting wire is induced by the current $I$ flowing in a driving conductor placed close to it. (b) Radiative drag in a set of particles:  a drag heat flux $J_d$  carried by thermal photons between two particles is induced by a heat fux $J$ exchanged between two thermostated particles  in a many body system.}
\end{figure}

To start, we consider a system as sketched in Fig.~1-b made with two parallel rows of nanoparticles  when the extremities of the first row are held at fixed temperature with two external thermostats while others particles relax to their own equilibrium temperature. 
Practically the particles can be, for instance, grafted at the end of elongated scanning probe microscope tips and they can be heated up by conduction through the tip itself with a resistive heater integrated to the tip.  The temperature of each particle can be controlled independently using those thermal resistors or left free to relax to their own equilibrium temperature.
In this system,  all particles can exchange electromagnetic energy between them as well as with their surrounding  which can be assimilated to a bosonic field at a given temperature $T_b$. 
(For subwavelength separation distances the power exchanged between the particles through near-field interactions is generally much more significant than the power exchanged with the surrounding bath so that the latter can be neglected). In steady state regime the unthermostated particles (i.e. $i\neq 1,N$) reach their equilibrium temperature $T_{ie}$. Using the Landauer theory in many body systems the  power exchanged between the $i^{th}$ and the $j^{th}$ particle in this network reads~\cite{Fan2,pba_Hall,Latella_PRL2017,Fan3}
\begin{equation}
  \varphi_{ij}=\int_{0}^{\infty}\frac{\rd\omega}{2\pi}\,[\Theta(\omega,T_{i})\mathcal{T}_{i,j}(\omega)-\Theta(\omega,T_{j})\mathcal{T}_{j,i}(\omega)]\label{Eq:InterpartHeatFlux},
\end{equation}
where $\Theta(\omega,T)={\hbar\omega}/[{e^{\frac{\hbar\omega}{k_B T}}-1}]$ is the mean energy of a harmonic oscillator in
thermal equilibrium at temperature $T$ and $\mathcal{T}_{i,j}(\omega)$ denotes the transmission coefficient, at the frequency $\omega$, between the two particles. When the particles are small enough 
compared with their thermal wavelength $\lambda_{T_{i}} = c\hbar/(\kb T_{i})$ ($c$ is the vacuum light
velocity, $2 \pi \hbar$ is Planck's constant, and $\kb$ is Boltzmann's constant) they can be modeled by simple radiating electrical dipoles.
In this case the transmission coefficient between the dipole $i$ and $j$ is defined as~\cite{Cuevas,Cuevas2}
\begin{equation}
\mathcal{T}_{i,j}(\omega)=\frac{4}{3}(\frac{\omega}{c})^4\Im\Tr\bigl[\hat{\boldsymbol\alpha}_j\mathds{G}_{ji}\frac{1}{2}(\boldsymbol{\alpha}_i-\boldsymbol{\alpha}_i^{\dagger})\mathds{G}_{ji}^{\dagger}\bigr],
\end{equation}
where $\hat{\boldsymbol\alpha}_i$ denotes the polarizability tensor of the $i^{th}$ particle and $\mathds{G}_{ij}$ is the dyadic Green tensor between the $i^{th}$ and the $j^{th}$ particle in the N-dipoles system~\cite{Purcell}.

\begin{figure}
\centering
\includegraphics[scale=0.34]{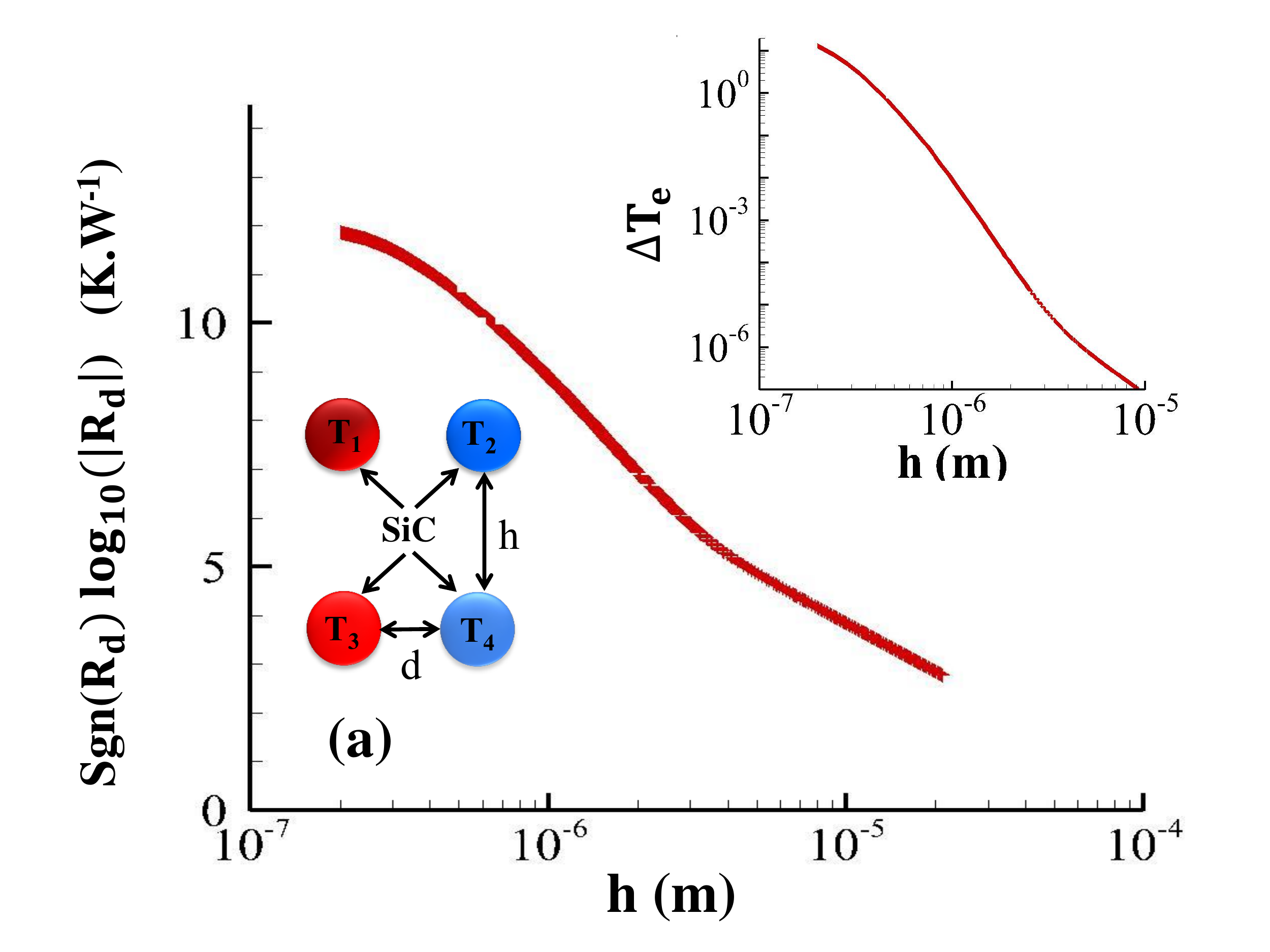}
\includegraphics[scale=0.34]{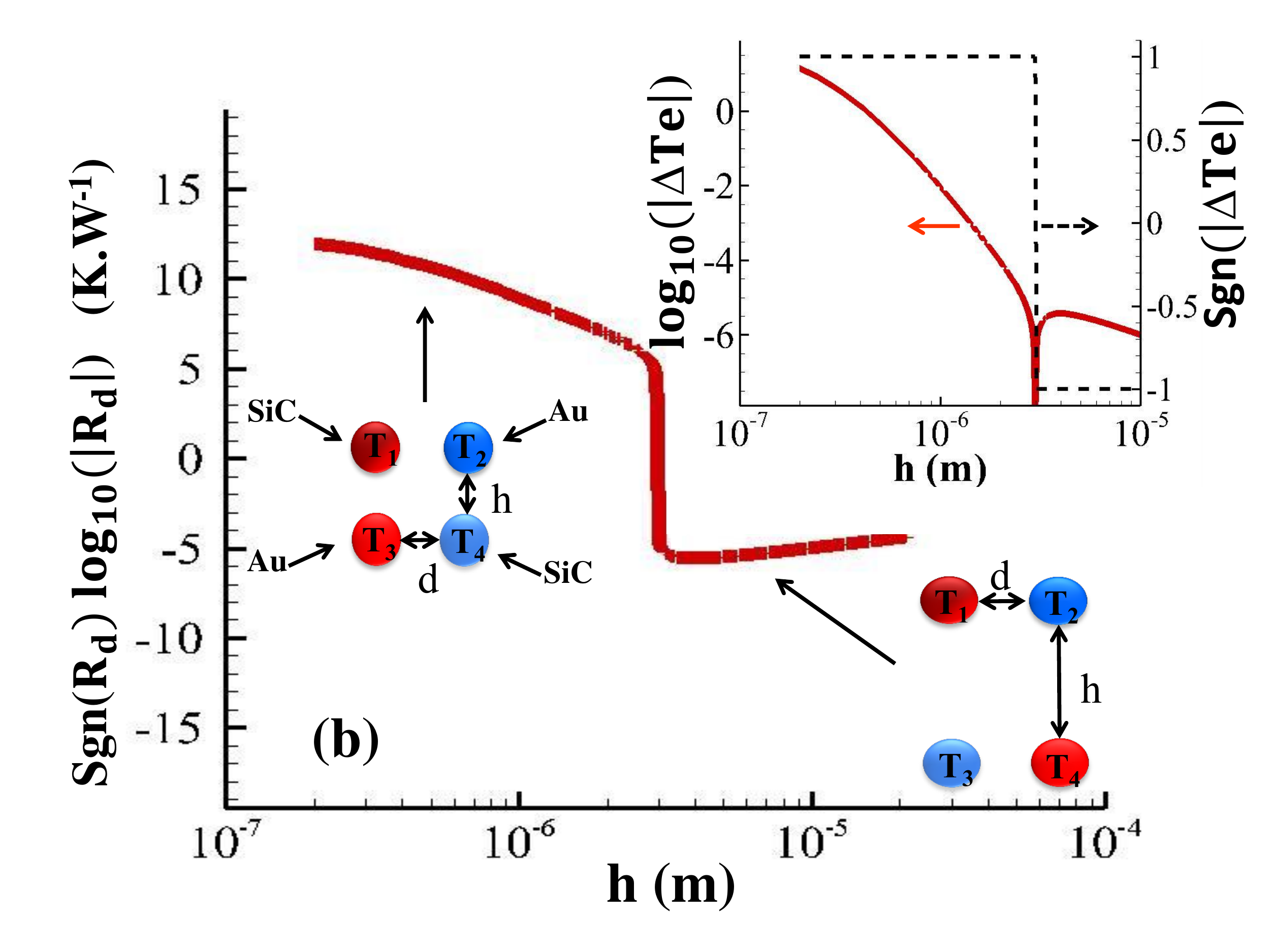}
\caption{Drag resistance between two dimers of (a)  SiC nanoparticles and (b) mixed  SiC/gold nanoparticles ($R=100 nm$) under a primary temperature difference $\Delta T=T_1-T_2=50$ when $T_2=300 K$ and $d=2R$ with respect to the dimers separation distance $h$.  Insets: temperature difference $\Delta T_e=T_3-T_4$ induced in the second dimer. Insets:temperature difference with respect to $h$. The dashed line in the inset of Fig. 2b shows the sign of the temperature difference.}
\end{figure}

From expression (\ref{Eq:InterpartHeatFlux}) the net thermal power received by each particle is simply given by summation over all incoming power (including the power coming from the thermostats) so that
\begin{equation}
  \phi_{i}=\underset{j\neq i}{\sum}\varphi_{ji}\label{Eq:HeatFlux}.
\end{equation}
Then, the net power exchanged  between the  first and the $N^{th}$ particle also called driving power reads
\begin{equation}
  J=\phi_{N}-\phi_{1}=\underset{j\neq N}{\sum}\varphi_{jN}-\underset{j\neq 1}{\sum}\varphi_{j1}\label{Eq:drivePower}.
\end{equation}
The calculation of these powers requires the knowledge of equilibrium temperatures $\bar{\mathbf{T}}=(T_{2e},...,T_{(N-1)e},T_{(N+1)e},...,T_{2Ne})^t$ which can be obtained from the energy balance equations at equilibrium
\begin{equation}
\begin{array}{c}
\phi_{th1}+\underset{j\neq 1}{\sum}\varphi_{j1}=0,\\
\phi_{thN}+\underset{j\neq N}{\sum}\varphi_{jN}=0,\\
\underset{i\neq (1,N)}{\underset{j\neq i}{\sum}}\varphi_{ji}=0,
\end{array}
\end{equation}
where $\phi_{th1}$ and  $\phi_{thN}$ stand for the power injected from the external thermostats into the first and the $N^{th}$ particles. Using  the thermal conductance $G_{ij}=\frac{\partial\varphi_{ij}}{\partial T}$ this system can be recasted into the following form
\begin{equation}
\begin{array}{c}
\phi_{th1}+\underset{j\neq 1}{\sum}G_{j1}(T_j-T_h)=0,\\
\phi_{thN}+\underset{j\neq N}{\sum}G_{jN}(T_j-T_c)=0,\\
\underset{i\neq (1,N)}{\underset{j\neq i}{\sum}}G_{ji}(T_j-T_i)=0.
\end{array}
\end{equation}
This set of equations also takes the form of following block system
\begin{equation}
\left(\begin{array}{ccc}
\mathbf{I}_{\mathbf{2}} & \vdots & \mathbf{A}\\
\cdots & \cdots & \cdots\\
\mathbf{0} & \vdots & \mathbf{C}
\end{array}\right)\left(\begin{array}{c}
\bar{\mathbf{\Phi}}\\
\bar{\mathbf{T}}
\end{array}\right)=\left(\begin{array}{c}
\mathbf{\bar{U}}\\
\mathbf{\bar{V}}
\end{array}\right)
\end{equation}
where $\mathcal{\mathbf{A}}=\left(\begin{array}{ccc}
G_{2,1} & \cdots & G_{2N,1}\\
G_{2,N} & \cdots & G_{2N,N}
\end{array}\right)$ and $\bar{\mathbf{U}}=(\underset{j\neq 1}{\sum}G_{j1}T_1,\underset{j\neq N}{\sum}G_{jN}T_N)^t$,
 $C_{ij}=-\underset{k\neq i}{\sum}G_{ki}\delta_{ij}+G_{ji}(1-\delta_{ij})$ with $i,j\neq(1,N)$
 and  $\bar{\mathbf{V}}=-((G_{1,2}T_1+G_{N,2}T_N),(G_{1,2N}T_1+G_{N,2N}T_N))^t$. In this system $\mathbf{I}_{\mathbf{2}}$ is the $2\times 2$ unit matrix and $\mathbf{0}=\left(\begin{array}{ccc}
0 & \cdots & 0\\
0 & \cdots & 0
\end{array}\right)^{t}$ is the zero matrix of format $(2N-2)\times2$ .
It turns out that the equilibrium temperatures reads
\begin{equation}
\bar{\mathbf{T}}=\mathcal{\mathbf{C}}^{-1}\bar{\mathbf{V}}\label{Eq:temperature},
\end{equation}
and the powers $\bar{\mathbf{\Phi}}=(\phi_{th1},\phi_{thN})^t$ coming from the two thermostats is simply given by the following expression
\begin{equation}
\bar{\mathbf{\Phi}}=\bar{\mathbf{U}}-\mathcal{\mathbf{A}}\bar{\mathbf{T}}\label{Eq:flux_thermostat}.
\end{equation}
Note that this reasoning can easily be extended to arbitrary many body systems with more than two thermostated particles.

Given the  power $J$ transmitted from particles 1 to N in the active layer the temperature difference $\Delta T_e=(T_{N+1,e}-T_{2N,e})$ induced in the passive layer is proportional to $J$ and  we can introduce the thermal drag resistance as
\begin{equation}
R_D=\frac{\Delta T_e}{J}\label{Eq:drag_resistance}
\end{equation}
which is a direct measure of many-body interactions between the two parallel layers.
In the particular case of two parallel dimers the equilibrium themperatures read
\begin{equation}
\bar{\mathbf{T}}=\frac{1}{\Delta}\left(\begin{array}{c}
\underset{j\neq4}{\Sigma}G_{j4}(G_{13}T_{1}+G_{23}T_{2})+G_{43}(G_{14}T_{1}+G_{24}T_{2})\\
\underset{j\neq3}{\Sigma}G_{j3}(G_{14}T_{1}+G_{24}T_{2})+G_{34}(G_{13}T_{1}+G_{23}T_{2})
\end{array}\right)
\end{equation}
with $\Delta={\displaystyle \underset{j\neq3}{\Sigma}G_{j3}\underset{j\neq4}{\Sigma}G_{j4}-G_{34}G_{43}}$. This leads after a straighforward calculation to the induced temperature difference between particles 3 and 4
\begin{equation}
T_3-T_4=\frac{G_{13} G_{24}-G_{14} G_{23}}{\Delta}(T_1-T_2).
\end{equation}
As for the primary current it writes
\begin{equation}
J=\underset{j\neq2}{\Sigma}G_{j2}(T_{j}-T_{2})-\underset{j\neq1}{\Sigma}G_{j1}(T_{j}-T_{1}),
\end{equation}
with $T_1=T_h$ and $T_2=T_c$ the temperatures set by the two thermostats. In Fig. 2-a we show the drag resistance between two dimers of silicon carbide (SiC) nanoparticles for different separation distances with respect to the primary temperature difference.  We describe the dielectric properties of SiC by means of a Drude-Lorentz model \cite{Palik98}
$\epsilon (\omega) =\epsilon_{\infty}\frac{\omega ^2-\omega_\text{L}^2+i\gamma\omega}{\omega^2-\omega_\text{R}^2+i\gamma\omega}$
with $\epsilon_{\infty}=6.7$, $\omega_{\rm L}=1.825\times 10^{14}\,$rad/s, $\omega_{\rm T}=1.494\times 10^{14}\,$rad/s and $\gamma=0.9\times 10^{12}\,$rad/s. As expected this resistance is positive and decays monototically with the separation distance $H$ between the two dimers. On the contrary   
in a mixed crossed system (Fig. 2-b) composed by SiC/gold nanoparticles (the dielectric permittivity of gold particles is given by the  Drude model $\epsilon (\omega) =1-\frac{\omega_p^2}{\omega(\omega+i\gamma)}$ with $\omega_p=13.71\times 10^{15}$ $rad.s^{-1}$ and $\gamma=4\times 10^{13} s^{-1}$~\cite{Palik98}) the situation radically changes and a negative drag effect appears in the intermediate regime between the near and far-field regimes. In this region the induced temperature difference $\Delta T_e$ becomes negative. As shown in Fig. 3 this negative drag effect is due to a better coupling along the diagonal between the two $SiC$ particles compared to  the vertical coupling between the SiC and the gold particles so that the $4^{th}$  particle is heated up furthermore than $3^{th}$ particle. This negative drag effect results from the spectral mismatch of SiC and gold particles.

It is worthwile to note that this drag effect also impacts the supercurrents which exist at equilibrium in some many-body systems at fixed temperature. As demonstrated by Fan et al.~\cite{Fan2} a  persistent directional heat current can arise in non-reciprocal networks.
In this case, when the many-body system is held at a given temperature $T$, it follows from the general expression (\ref{Eq:InterpartHeatFlux}) that the body $i$ and $j$ in this system still exchange an energy flux~\cite{Latella_PRL2017}
\begin{equation}
\varphi^{eq}_{ij}=\int_{0}^{\infty}\frac{\rd\omega}{2\pi}\,\Theta(\omega,T)[\mathcal{T}_{i,j}(\omega,\mathbf{B})-\mathcal{T}_{j,i}(\omega,\mathbf{B})]\label{Eq:equilibriumHeatFlux}.
\end{equation}
Notice that while the non-equilibrium flux given by expression (\ref{Eq:InterpartHeatFlux}) depends on the temperature difference between the particles $i$ and $j$, this flux is related to the optical non-reciprocity within the system and it can exist even at thermal equilibrum.

\begin{figure}
\centering
\includegraphics[scale=0.3]{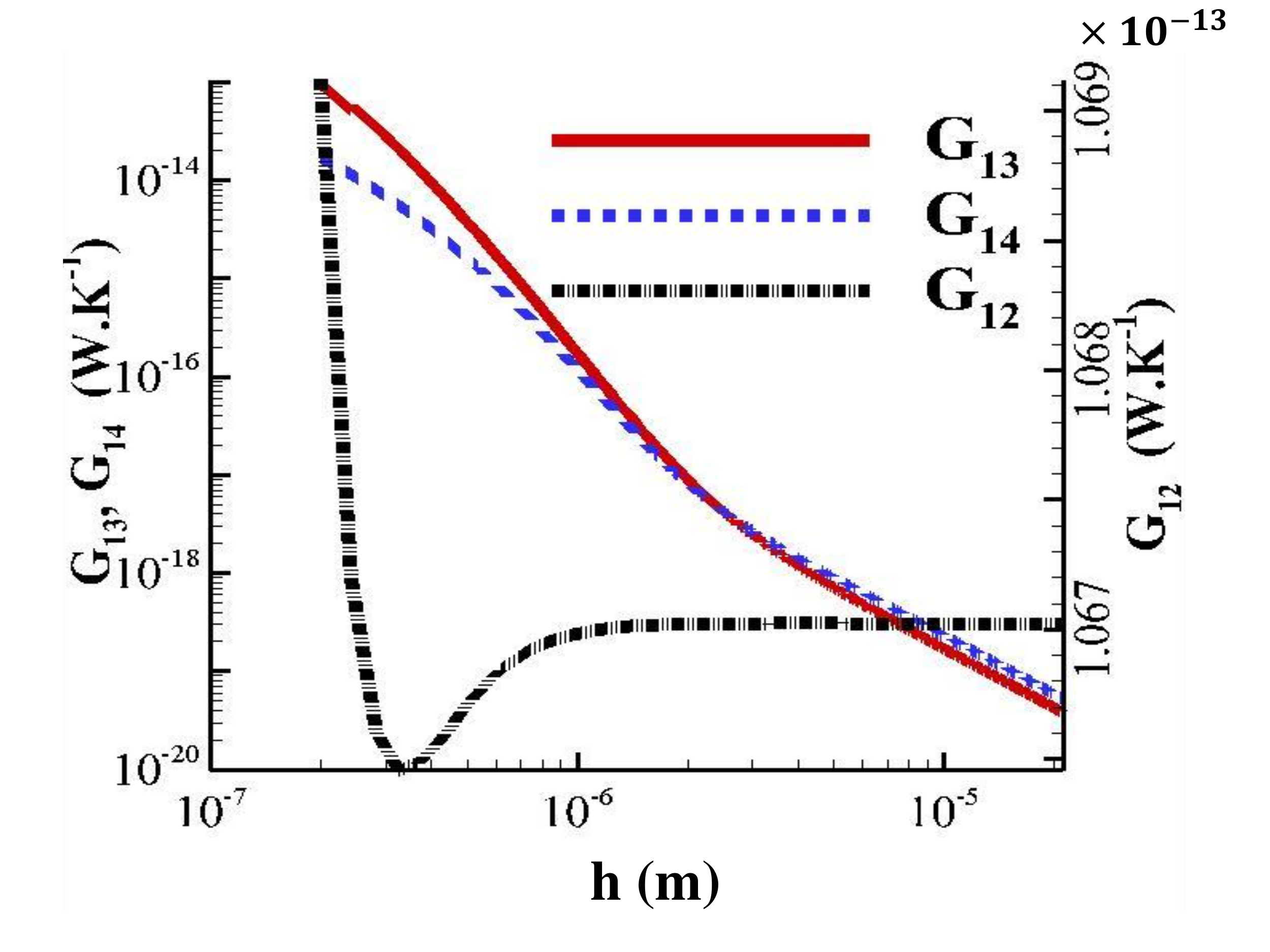}
\caption{Thermal conductance at $T=300 K$ between the particles in the same system as in Fig. 2-b  with respect to the dimers separation distance$h$ between the two dimers.}
\end{figure}

To illustrate how this persistent heat current is affected by the frictional effect let us consider a system made with two parallel magneto-optical networks of regularly distributed InSb nanoparticles along  parallel rings as sketched in (Fig. 4) and submitted to an external magnetic field applied perpendicularly to these rings.  The dielectric tensor associated to  InSb particles takes the following form~\cite{Palik,Moncada}
\begin{equation}
\bar{\bar{\varepsilon}}=\left(\begin{array}{ccc}
\varepsilon_{1} & -i\varepsilon_{2} & 0\\
i\varepsilon_{2} & \varepsilon_{1} & 0\\
0 & 0 & \varepsilon_{3}
\end{array}\right)\label{Eq:permittivity}
\end{equation}
with
\begin{equation}
\varepsilon_{1}(B)=\varepsilon_\infty(1+\frac{\omega_L^2-\omega_T^2}{\omega_T^2-\omega^2-i\Gamma\omega}+\frac{\omega_p^2(\omega+i\gamma)}{\omega[\omega_c^2-(\omega+i\gamma)^2]}) \label{Eq:permittivity1},
\end{equation}
\begin{equation}
\varepsilon_{2}(B)=\frac{\varepsilon_\infty\omega_p^2\omega_c}{\omega[(\omega+i\gamma)^2-\omega_c^2]}\label{Eq:permittivity2},
\end{equation}
\begin{equation}
\varepsilon_{3}=\varepsilon_\infty(1+\frac{\omega_L^2-\omega_T^2}{\omega_T^2-\omega^2-i\Gamma\omega}-\frac{\omega_p^2}{\omega(\omega+i\gamma)})\label{Eq:permittivity3}.
\end{equation}
Here, $\varepsilon_\infty=15.7$ is the infinite-frequency dielectric constant, $\omega_L=3.62\times10^{13} rad.s^{-1}$ is the longitudinal optical phonon frequency, $\omega_T=3.39\times10^{13} rad.s^{-1}$ is the transverse optical phonon frequency, $\omega_p=(\frac{ne^2}{m^*\varepsilon_0\varepsilon_\infty})^{1/2}$ is the plasma frequency of free carriers of density $n=1.36\times10^{19} cm^{-3}$ and effective mass $m^*=7.29\times 10^{-32}kg$, $\Gamma=5.65\times10^{11} rad.s^{-1}$ is the phonon damping constant,$\gamma=10^{12} rad.s^{-1}$ is the free carrier damping constant and $\omega_c=eB/m^*$ is the cyclotron frequency. Thus, the polarizability tensor for a spherical particle can be described, including the radiative corrections, by the following anisotropic polarizability~\cite{Albaladejo}
\begin{equation}
\bar{\bar{\boldsymbol{\alpha}}}_{i}(\omega)=( \bar{\bar{\boldsymbol{1}}}-i\frac{k^3}{6\pi} \bar{\bar{\boldsymbol{\alpha_0}}}_{i})^{-1} \bar{\bar{\boldsymbol{\alpha_0}}}_{i}\label{Eq:Polarizability},
\end{equation}
where $ \bar{\bar{\boldsymbol{\alpha_0}}}_{i}$ denotes the quasistatic polarizability of the $i^{th}$ particle which reads for spheres of radius $R$ in vacuum $\varepsilon_h$
\begin{equation}
  \bar{\bar{\boldsymbol{\alpha_0}}}_{i}(\omega)=4\pi R^3(\bar{\bar{\varepsilon}}-\bar{\bar{1}})(\bar{\bar{\varepsilon}}+2\bar{\bar{1}})^{-1}\label{Eq:Polarizability2}.
\end{equation}

\begin{figure}
\centering
\includegraphics[scale=0.35]{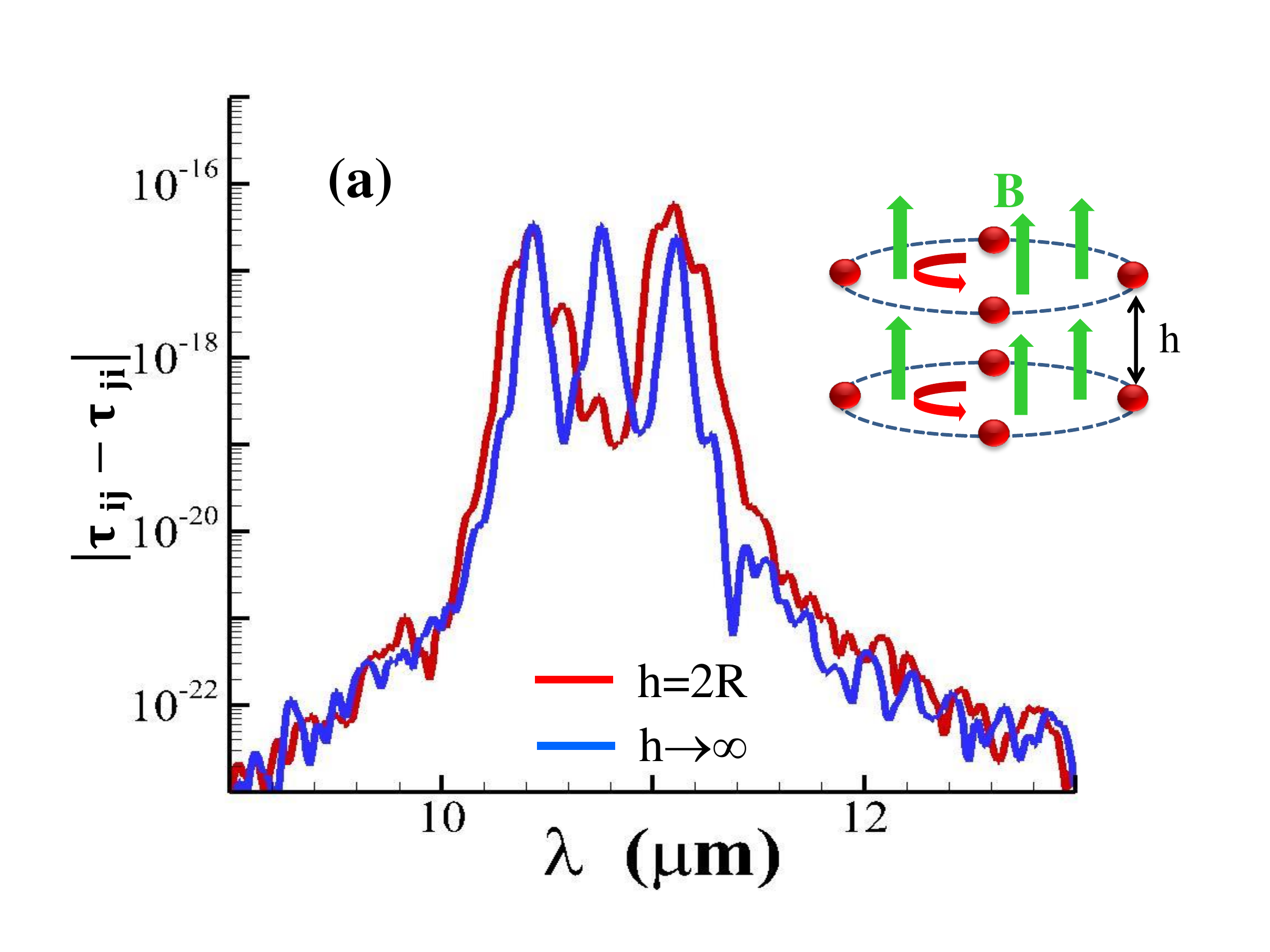}
\includegraphics[scale=0.35]{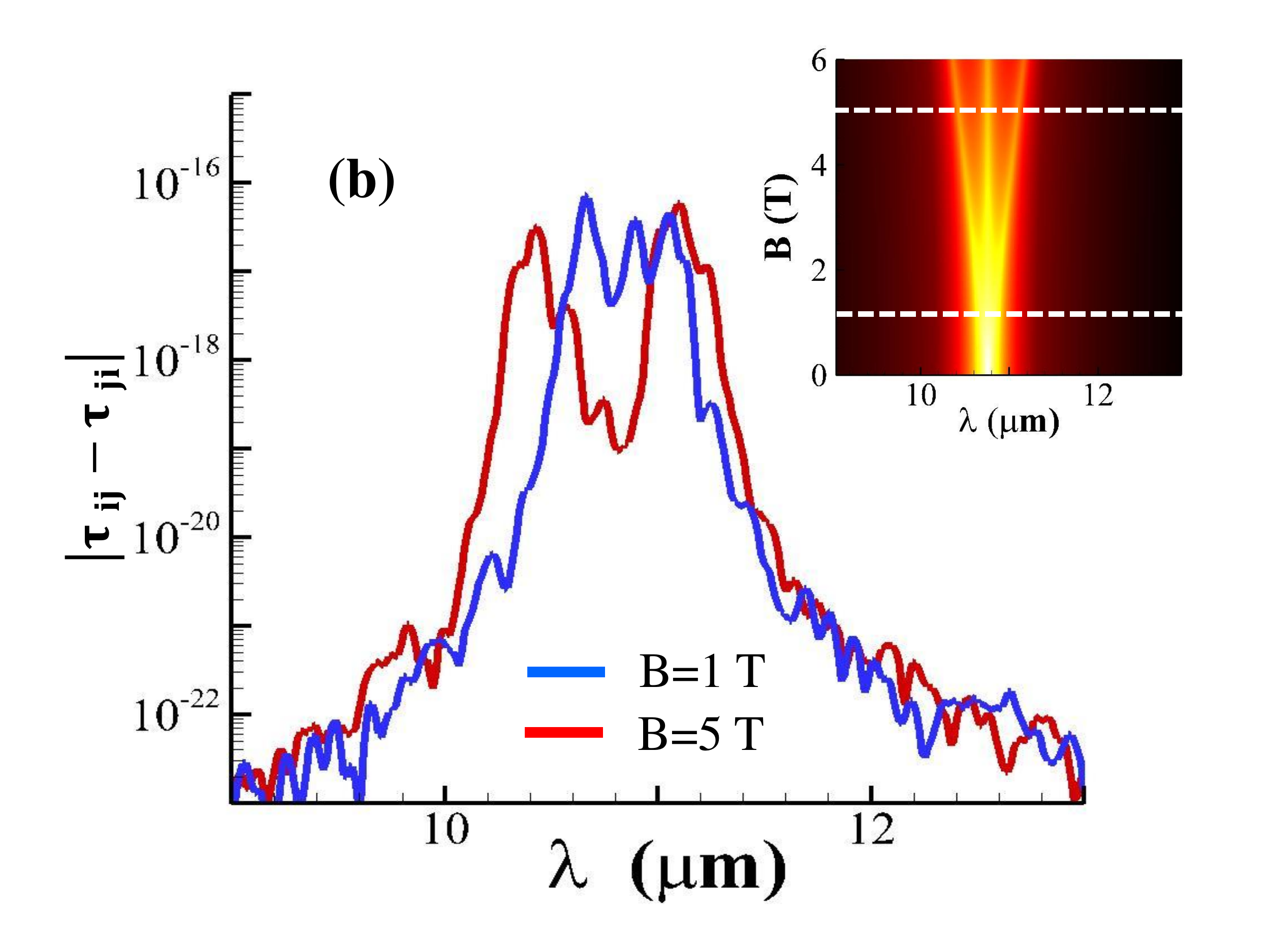}
\caption{Heat supercurrent spectrum (normalized by the energy of Planck oscillator)  in  a  many-body system of 8 InSb particles ($R=100 nm$) regularly and identically distributed on two parallel rings under an external magnetic field $B$  orthogonal to the rings for (a) two separation distances with $B=5\: T$ and for (b) two magnetic field intensities when $h=2\: R$. Inset of (b):  resonances of an isolated InSb particle shown in the $(\lambda,B)$ plane.}
\end{figure}

In Fig. 4-a we show how the supercurrent is  modified with respect to the separation distance between the rings under the action of an external magnetic field of $5$ Tesla. At large separation distance we see that the supercurrent spectrum displays three distinct resonant peaks in the midinfrared around $11 \mu m$. Those peaks correspond to the resonances of isolated InSb particle~\cite{Latella_PRL2017} as shown in the inset of Fig.4-b. When the separation distance between the two rings is reduced to ($h=2\:R$) we observe that the spectrum is broadened toward the higher wavelengths meaning so that the supercurrent can be amplified by the interplay of two networks when the Wien's wavelength of system is in this spectral range. On the contrary if the equilibrium temperature is sufficiently increased  the Planck window where energy exchanges occur is shifted toward the lower wavelengths and the supercurrent is inhibated.
This broadening simply results from the increase in the number of coupling channels between the two networks. In Fig.4-b we show that the drag effect on the persistent current also is sensitive to the magnitude of magnetic field or in other words to the importance of the optical non-reciprocity inside the system. When $B=1\:T$ the supercurrent spectrum is relatively narrow and centered around the resonant wavelengths of each InSb particle (see Inset of Fig. 4-b). However, comparerd to what happens at  $B=5\:T$, in this case the splitting of resonances is less pronounced.  As the magntitude of magnetic field increases the non-reciprocity in the system becomes more important and the splitting between the resonant modes of InSb particles become more signitficant.  It turns out that the supercurrent specturm widens and this widening mainly occurs at low wavelengths.
Finally I emphasize in this section the differences which exist between the supercurrents interaction and the coupling between magnetic dipoles produced for instance by loops of electric current or by free charges oscillation in split-ring resonators.  Let us consider for clarity reasons the case of two circular  loops of same area $S$ which carry a primary current $I(t)$. Then this current generates a magnetic moment $\bold{m}=I.\bold{S}$ ($\bold S$ being the oriented surface of the loop) which gives rise to an electromagnetic field in its surrounding environment. Then, according to Faraday's law this field leads to an electromotive force in the neighboring loop which in turn induced an electric current $I_d$.  In the particular case of  two parallel loops this induced current is related to $I$ by the simple relation $I_d=\frac{M}{R}\frac{dI}{dt}$ where $M$ denotes the mutual inductance and $R$ the electric resistance inside each loop ( $M=\frac{\mu_0}{2\pi d^3 }\frac{S^2}{R}$ ,  $d$ being their separation distance and $\mu_0$ the vacuum permeability). On the contrary, in the thermal problem the persistent current is a flow of energy which is carried by the electromagnetic field itself and not by electric charges. This field weakly impacts the cyclotronic motion within the magneto-optical material so that it does not induced an electromotive force. The interplay between the two systems is only due to the electric dipoles coupling. Nevertheless  the circular heat flux is connected to an orbital angular momentum  of the electromagnetic field and also to a spin angular momentum~\cite{Bliokh}. The consequences of these quantities and of their mutual interaction in the system environment  could be used to detect the existence of persistent heat flux in a similar way as the persistent electric currents predicted by the quantum mechanics~\cite{Buttiker,Bleszynski}. 

In summary, we have introduced a drag effect induced by the electromagnetic interactions in many-body systems. This frictional effect quantifies the strength of electromagnetic  interactions in these systems and it gives insights on the spatial distribution of temperature in non-equilibrium situations. By  introducing the concept of drag resistance for the thermal photons we have demonstrated the existence of regions in these systems where the heat flux can locally flows in an opposite direction to the local temperature variation. 
Beside we have shown that this drag effect also exists at thermal equilibrium between persistent heat flux. By investigating the frictional drag between two parallel circular loops of magneto-optical particles under the action of a magnetic field we have demonstrated that the heat superccurents supported by these systems can either be  inhibitated or amplified by this effect. Furthermore, we think that  the  angular momentum and spin associated to these persistent heat flux could be use to conceive an experimental setup in order to prove their existence.

\begin{acknowledgements}
%P.B.-A. acknowledges discussions with .
\end{acknowledgements}

\end{document}